# Data Assimilation of Satellite Fire Detection in Coupled Atmosphere-Fire Simulation by WRF-SFIRE

Jan Mandel[a], Adam K. Kochanski[b], Martin Vejmelka[c],
and Jonathan D. Beezley[d]

[a]*University of Colorado Denver, Denver, CO, USA, jan.mandel@gmail.com*
[b]*University of Utah, Salt Lake City, UT, USA, adam.kochanski@utah.edu*
[c]*University of Colorado Denver, Denver, CO, USA, and Institute of Computer Science, Academy of Sciences of the Czech Republic, Prague, Czech Republic, vejmelkam@gmail.com*
[d]*Kitware, Inc., Clifton Park, NY, USA, jon.beezley@gmail.com*

**Abstract**
Currently available satellite active fire detection products from the VIIRS and MODIS instruments on polar-orbiting satellites produce detection squares in arbitrary locations. There is no global fire/no fire map, no detection under cloud cover, false negatives are common, and the detection squares are much coarser than the resolution of a fire behavior model. Consequently, current active fire satellite detection products should be used to improve fire modeling in a statistical sense only, rather than as a direct input. We describe a new data assimilation method for active fire detection, based on a modification of the fire arrival time to simultaneously minimize the difference from the forecast fire arrival time and maximize the likelihood of the fire detection data. This method is inspired by contour detection methods used in computer vision, and it can be cast as a Bayesian inverse problem technique, or a generalized Tikhonov regularization. After the new fire arrival time on the whole simulation domain is found, the model can be re-run from a time in the past using the new fire arrival time to generate the heat fluxes and to spin up the atmospheric model until the satellite overpass time, when the coupled simulation continues from the modified state.

## 1. Introduction

Active fire detection products using the VIIRS and MODIS instruments provide planet-wide monitoring of fire activity several times daily as detection squares or polygons at a resolution of 375 m to 1 km. Because the data products are continuously available online, they present an attractive data source for automated fire behavior simulations and forecasts. Unfortunately, fire detection errors are frequent (Csiszar et al., 2012; Hawbaker et al., 2008; Sei, 2011) , and geolocation errors can be significant, up to 1.5 km (Sei, 2011). An improved data product at 375 m resolution and with better error rates exists (Schroeder et al., 2014), but it is not available for general use yet. In any case, satellite active fire detection data have significant error rates, and they have much coarser resolution than the resolution of fire behavior models, which is typically few meters to tens of meters.

Fire detection data consist of squares of polygons (from now on, squares) with associated satellite overpass times. A fire detection square means that a fire of sufficient intensity and size was detected somewhere in the square; it does not mean that the whole square is burning. The fire detection squares are in arbitrary locations; a global fire-detection map where every pixel is marked as either fire or no fire is "neither required or desired" by VIIRS Active Fires specifications (Sei, 2011). In addition, an existing fire is often not detected (a false negative). For example, MODIS has a 50% probability of detecting a $100m^2$ flaming fire (Giglio et al., 2003). The improved 375m

VIIRS product has 50% probability of detecting about a 2 m$^2$ flaming fire at night, and about 40 m$^2$ flaming fire during the day of (Schroeder et al., 2014). False positives are possible, particularly in areas of high contrast (the data products' algorithms are contextual), but less frequent. Finally, if a location is under cloud cover, the fire detection is turned off completely. Unfortunately, the cloud mask used was not provided in the data products we found, so we could not distinguish between missing data and negative fire detection.

Consequently, current satellite fire detection products do not determine the fire area and the fire perimeter to a degree that could be relied upon. Although the use of satellite fire detection to initialize a fire simulation directly was an important advance (Coen and Schroeder, 2013), satellite fire detection is better suited to improve a fire behavior simulation in a statistical sense only, that is, by data assimilation. Data assimilation fuses the forecast obtained from a model with the data by balancing their uncertainties, and it can also take advantage of more reliable fire detection in future by putting more weight on the data and less weight on the model.

Data assimilation modifies the state of the simulation in analysis cycles. Each cycle consists of advancing the model in time and an analysis step, which takes account of the new information. The analysis cycles steer the simulation periodically and help to avoid an accumulation of modeling errors. Analysis steps at every satellite overpass also help to account for uncertainties caused by incorrect fuel information and by fire-fighting efforts. We propose a new data assimilation method for satellite fire detection, inspired by techniques used for contour detection in computer vision, such as in the Microsoft Kinect (Blake, 2014). The method takes advantage of encoding the state of the fire propagation as the fire arrival time (Finney, 2002), which can be modified by an additive correction. We employ a Bayesian approach and obtain a Maximum-a-Posteriori (MAP) estimate as a solution of a generalized nonlinear least squares problem, similarly as in Stuart (2010), to simultaneously maximize the log likelihood of the fire detection (or lack od detection) and minimize the change in the fire arrival time. The method is implemented efficiently using Fast Fourier Transform (FFT), and its computational cost is negligible compared to the coupled atmosphere-fire simulation itself.

Another challenge of data assimilation in coupled fire-atmosphere models is how to change the state of the atmospheric model when the state of the fire model changes in response to data. Atmospheric circulation evolves in response to the heat flux from the fire over time, and when the state of the fire model changes, the consistence between the state of fire and the state of the atmosphere is lost. Encoding the fire model state as the fire arrival time allows to go back in time, blend the new fire arrival time with the original one, and spin up the atmospheric model with heat fluxes generated by replaying the modified fire arrival time instead of running the fire propagation model itself (Mandel et al., 2012). Then, at the fire detection time (the satellite overpass time), the fire spread model takes over, and the simulation continues.

## 2. Methods

### 2.1. Fire detection data likelihood

We assimilate the fire detection data in the form of a likelihood of the detection data at a location $(x, y)$. The data likelihood is proportional to $e^{f(t,x,y)}$, where $t$ is the number of hours the fire has arrived at the location $(x, y)$ before the satellite overpass time. The function $f(t,x,y)$ is called log likelihood. For locations $(x, y)$ within the fire detection squares, the data likelihood should be high for when the fire has arrived recently, and low otherwise. For locations outside of the fire detection squares, the data likelihood should be high if the fire has not arrived yet, or has arrived long ago, and low if the fire has arrived at the location recently. Of course, this is a very simplified view; more accurately, the fire detection data likelihood depends on the fire behavior (some fires burn longer and can be detected longer after the fire arrival, and some only for a shorter time), the atmospheric state between the fire location and the satellite, and the properties of the sensor and processing algorithms. Some active fire detection products also provide a level of confidence, e.g. Schroeder et al. (2014), which should be a part of the calculation of the data likelihood. In this paper, however, we consider only a simple data likelihood functions modelled as in Fig. 1.

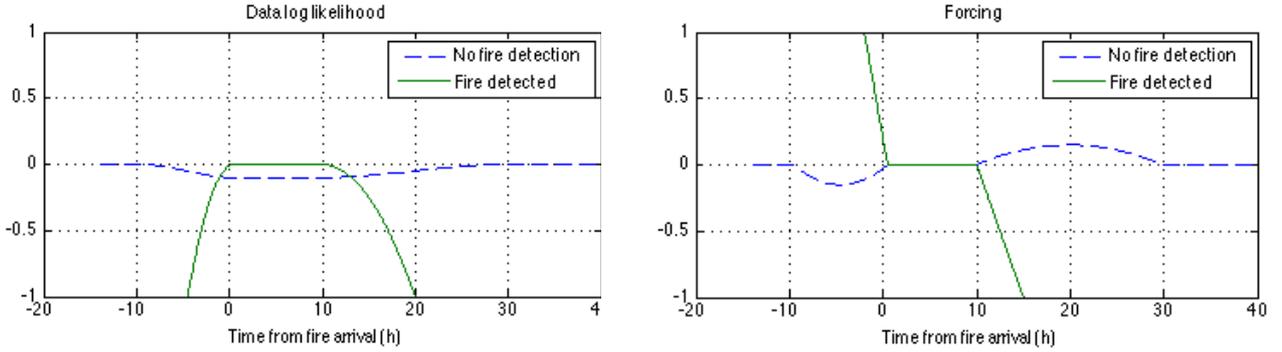

*Figure 1 - The log likelihood function $f(t)$ (left) and its derivative $f'(t)$ (right) as a function of time $t$ elapsed since fire arrival (hours), in locations inside and outside of fire detection squares.*

The model log likelihood $f$ shown in Fig. 1 is a function with continuous first derivative in the time $t$ since fire arrival, and it is specified by parameters $T_{\min}$, $T_{\max}$, $T_{\text{pos}}$, $T_{\text{neg}}$, $\psi > 0$. Within the fire detection squares, we choose $f(t) = 0$ for $T_{\min} \leq t \leq T_{\max}$, the function $f$ is a quadratic polynomial for $t < T_{\min}$ with $f(T_{\min} - T_{\text{neg}}) = -1$, and a quadratic polynomial for $t > T_{\max}$ with $f(T_{\max} + T_{\text{pos}}) = -1$. Outside of the fire detection squares, $f(t) = -\psi$ has a constant negative value for $T_{\min} \leq t \leq T_{\max}$, $f(t) = 0$ for $t < T_{\min} - 2T_{\text{neg}}$ or $t > T_{\max} + 2T_{\text{pos}}$, and $f$ is blended by cubic polynomials in between. However, if the location is under a cloud cover, we would use $f(t) = 0$ for any $t$.

For the sake of simplicity, the data likelihoods at different locations are assumed to be independent. Then, the overall data likelihood of all detection squares at the satellite overpass time $T^{\text{S}}$ over the whole simulation domain, given fire arrival time $T = T(x,y)$, is proportional to the product of the data likelihoods at all grid nodes $(x,y)$. We also weigh each log likehood by the cell area $\Delta x \Delta y$, which gives

$$p(\text{detection squares} \,|\, T) \propto \prod_{(x,y)} e^{f(T^{\text{S}}-T,x,y)\Delta x \Delta y} = e^{\sum_{(x,y)} f(T^{\text{S}}-T,x,y)\Delta x \Delta y} \approx e^{\int f(T^{\text{S}}-T,x,y)dx\,dy}, \qquad (1)$$

with the understanding that the integral is computed numerically over the fire simulation domain.

**2.2. Analysis step**

In the analysis step, the data likelihood (1) is combined with the forecast fire arrival time $T^{\text{f}} = T^{\text{f}}(x,y)$ to obtain the analysis fire arrival time $T^{\text{a}}$. For this purpose, we need the probability density of the forecast fire arrival time. We assume that the difference of the fire arrival time from the mean fire arrival time $T^{\text{f}}$ is a smooth random function, and we model the probability density of the fire arrival time as Gaussian, with mean $T^{\text{f}}$ and covariance $A^{-1}$ such that the associated norm

$$\|u\|_A = \langle u, Au \rangle^{1/2}$$

is small for spatially smooth functions $u$, defined on the model grid, and large for oscillatory $u$. For this purpose, we choose the covariance $A^{-1}$ as a negative power of a discretization of the Laplace operator

$$\Delta u = \frac{\partial^2 u}{\partial x^2} + \frac{\partial^2 u}{\partial y^2}$$

on the model grid, multiplied by a penalty parameter. Thus, the forecast probability density of the fire arrival time $T$ is

$$p^{\mathrm{f}}(T) \propto e^{-\frac{\alpha}{2}\|T-T^{\mathrm{f}}\|_A^2}, \quad A = (-\Delta)^p, \quad \alpha, p > 0.$$

We also requite that the ignition time of the fire does not change, so $p^{\mathrm{f}}(T) > 0$ only if $T = T^{\mathrm{f}}$ at the ignition point. The analysis probability density is obtained from the Bayes theorem as

$$p^{\mathrm{a}}(T) \propto p(\text{detection squares}|T) p^{\mathrm{f}}(T) \propto e^{\int f(T^{\mathrm{S}}-T,x,y)dx\,dy} e^{-\frac{\alpha}{2}\|T-T^{\mathrm{f}}\|_A^2} = e^{\int f(T^{\mathrm{S}}-T,x,y)dx\,dy - \frac{\alpha}{2}\|T-T^{\mathrm{f}}\|_A^2}.$$

The analysis distribution is non-Gaussian due to the presence of the data likelihood. To obtain a single analysis value, we use the standard Maximum-a-Posteriori estimator, which is the maximizer $T^{\mathrm{a}}$ of the analysis probability density $p^{\mathrm{a}}(T)$. Maximization of the analysis density is equivalent to the minimization of the exponent, which is the generalized least squares problem

$$J(T) = \frac{\varepsilon}{2}\|T-T^{\mathrm{f}}\|_A^2 - \int f(T^{\mathrm{S}}-T,x,y)\,dx\,dy \to \min_{T:\,T=T^{\mathrm{f}}\text{ at the ignition point}}. \quad (2)$$

Note that the optimization problem (2) can be also understood as the maximization of the log likelihood with a Tikhonov regularization by the added quadratic term. In this context, the matrix $A$ is chosen to penalize non-smooth increments $T - T^{\mathrm{f}}$. To find a descent direction, consider the difference

$$J(T+h) - J(T) = \frac{\alpha}{2}\langle A(T-T^{\mathrm{f}}+h), T-T^{\mathrm{f}}+h\rangle - \int f(T^{\mathrm{S}}-T-h,x,y)\,dx\,dy$$
$$- \left(\frac{\alpha}{2}\langle A(T-T^{\mathrm{f}}), T-T^{\mathrm{f}}\rangle - \int f(T^{\mathrm{S}}-T,x,y)\,dx\,dy\right)$$
$$= \alpha\langle A(T-T^{\mathrm{f}}), h\rangle + \int \langle \frac{\partial}{\partial t} f(T^{\mathrm{S}}-T,x,y)dx\,dy, h\rangle + O(\|h\|^2)$$
$$= \langle \nabla J(T), h\rangle + O(\|h\|^2),$$

which gives the gradient of $J$ as

$$\nabla J(T) = \varepsilon A(T-T^{\mathrm{f}}) - F(T), \text{ where } F(T) = \int \frac{\partial}{\partial t} f(T^{\mathrm{S}}-T,x,y)\,dx\,dy$$

is from now on called the forcing function. This gradient direction, however, is not suitable for our purposes, because it is not very spatially smooth, and, consequently, gradient descent iterations converge to the smoother solution of (2) only very slowly and, in addition, they are liable to get caught in local minima. A better gradient direction is obtained by considering the gradient $\nabla^A J(T)$ with respect to the inner product $\langle u,v\rangle_A = \langle Au,v\rangle$:

$$\langle \nabla^A J(T), h\rangle_A = \langle A\nabla^A J(T), h\rangle = \langle \nabla J(T), h\rangle$$
$$\nabla^A J(T) = \alpha(T-T^{\mathrm{f}}) - A^{-1} F(T).$$

Now we add the constraint that the increment $h$ is zero at the ignition point and introduce Lagrange multiplier $\lambda$, which leads to the saddle point problem

$$\varepsilon Ah + C\lambda = \alpha A(T^{\mathrm{f}}-T) + F(T)$$
$$C^{\mathrm{T}} h = 0, \quad (3)$$

for a descent direction $h$, where $C$ is a zero-one vector such that $C^{\mathrm{T}} h = 0$ expresses the condition that $h$ is zero at the ignition point. The saddle point problem (3) is easily solved with access to multiplication by $A^{-1}$ only: Calculating $h$ from the first equation in (3) gives the descent direction

$$h = T^{\mathrm{f}} - T + \alpha^{-1} A^{-1} F(T) - \alpha^{-1} A^{-1} C\lambda = -\alpha^{-1}\left(\nabla^A J(T) + A^{-1} C\lambda\right),$$

where the Lagrange multiplier is determined by substituting in the second equation, which gives

$$\lambda = \left(C^{T} A^{-1} C\right)^{-1} C^{T} \left(\alpha(T^{f} - T) + A^{-1} F(T)\right) = \left(C^{T} A^{-1} C\right)^{-1} C^{T} \nabla^{A} J(T).$$

Note that the descent direction $h$ is obtained by a spatial smoothing of the forcing $F(T)$, or, equivalently, from the gradient in the inner product defined by the inverse covariance $A$, plus an adjustment for the constraint that the fire arrival time at the ignition location should not change. In each step of the descent method, the cost function $J$ is then minimized by a line search over $T + h\tau$, $\tau$ real. The first iteration starts from $T = T^{f}$. Note that the penalty parameter $\alpha$ does not have any effect on the first search direction.

## 3. Results

The method is illustrated on two real fires simulated by WRF-SFIRE, with the assimilation of MODIS and VIIRS Active Fire detection data. WRF-SFIRE (Mandel et al., 2009, 2011) builds on CAWFE (Clark et al., 2004), and it is available from [openwfm.org](openwfm.org). A limited version from 2010 is currently distributed in WRF release as WRF-Fire (Coen et al., 2013).

In both examples below, we have used the parameters

$$\alpha = 1000, T_{min} = 0.5, T_{max} = 10, T_{pos} = 10, T_{neg} = 5, \psi = 0.5, p = 1.02.$$

### 3.1. 2012 Wood Hollow fire

The present data assimilation method was applied to adjust the fire arrival time obtained from a fire simulation up to 24$^{th}$ Jan 2014 23:45 UTC, from 151 MODIS fire detections, which occurred on the same day approximately at 20:00 UTC (Fig. 2). The practical implementation of the method implements a line search strategy, in which a maximal step is set (here, 1.0) and the objective function is sampled 6 times in regular intervals. The step with the lowest objective function value and its two neighbors are selected as the new limits and the line search is repeated to determine the best step size with more precision.

In this instance, only the first iteration resulted in an improvement of the objective function, a line search along the second direction did not yield any more improvements, even for smaller step sizes. However, the single iteration has already found a satisfactory modification of the fire arrival time. The seemingly large value of the penalty parameter $\alpha$ was needed to stop the method from adjusting the fire arrival time in later iterations near the fire detection squares only.

### 3.2. 2012 Barker Canyon fire

The data assimilation method was applied to a simulation of the 2012 Barker Canyon Fire, 1.71 days from ignition. Figs. 3-6 show how the analysis fire arrival time is obtained from the forecast by adding a correction in the direction of a smooth search direction, which is in turn obtained by the spatial smoothing of the forcing function. Again, a single iteration was sufficient to find a satisfactory solution.

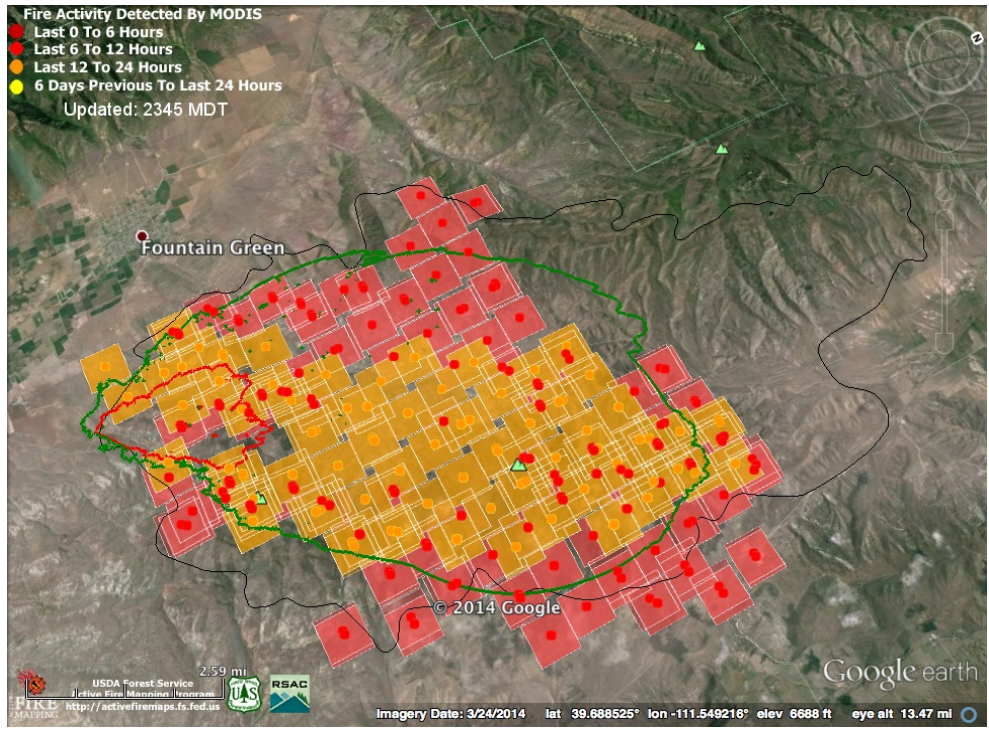

*Figure 2 – Data assimilation for the 2012 Woods Hollow fire, simulated by WRF-SFIRE, with MODIS Active Fire detection squares. The black contour is the actual fire perimeter on 25th Jun 2012, the red line is the forecast fire perimeter and the green line is the fire perimeter after assimilation, or the analysis. The analysis adjusts the shape of the fire for the fire detection data by a smooth correction, which automatically fills the fire arrival time*

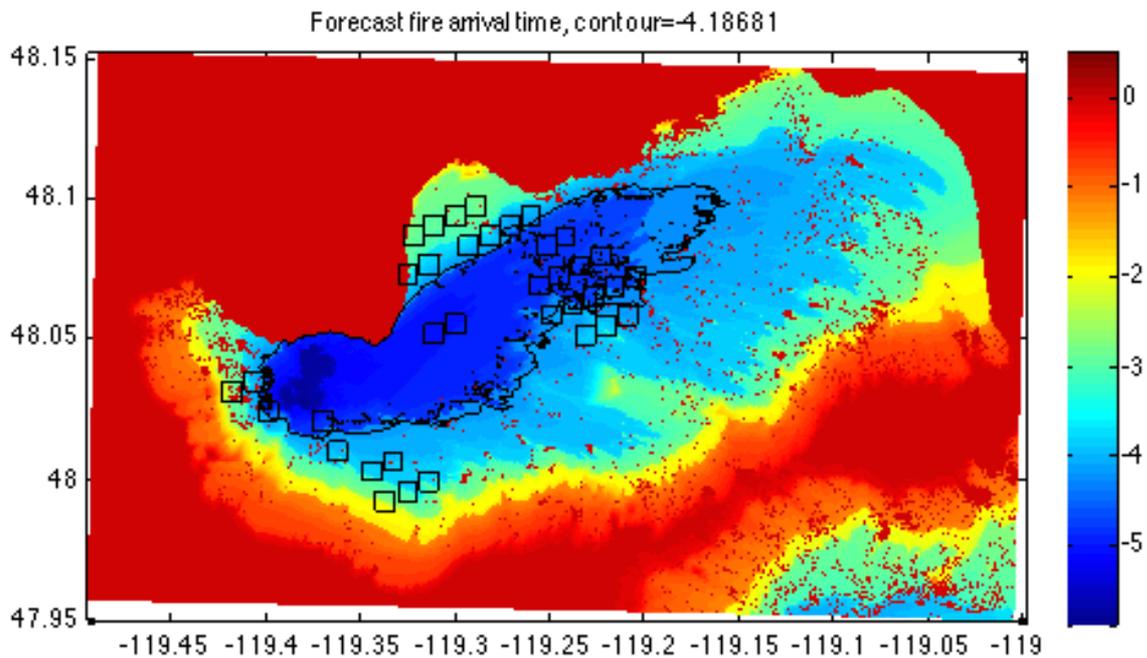

*between the detection squares.*

*Figure 3 – Forecast fire arrival time for the 2012 Barker fire, simulated by WRF-SFIRE, and VIIRS Active Fire detection squares. The black contour is the simulated fire perimeter at the satellite overpass time. The color bar shows fire arrival time in days before the end of the simulation. The axes are longitude and latitude.*

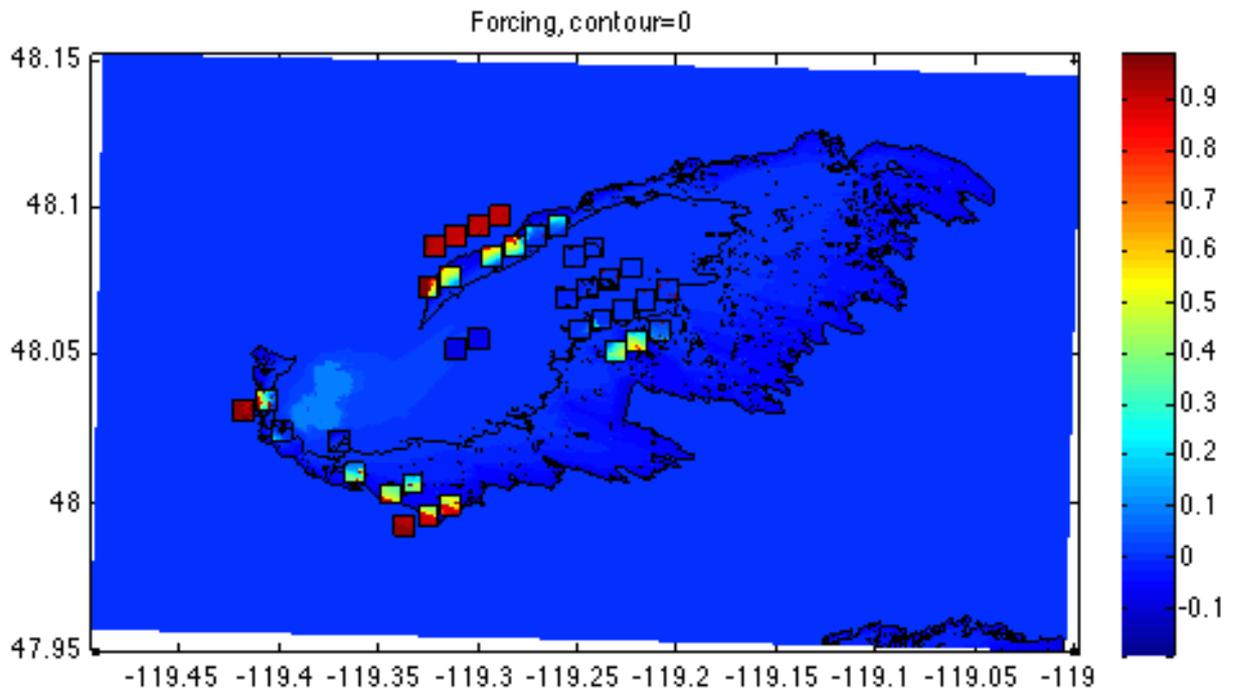

*Figure 4 – Forcing function F for the forecast in Fig. 3. Positive values mean that the fire arrival time at the location should be adjusted down, while negative value mean that the fire arrival time should be adjusted up. The black contour marks the locations where the forcing is zero. Note that the forcing has jumps at the boundaries of the detection squares.*

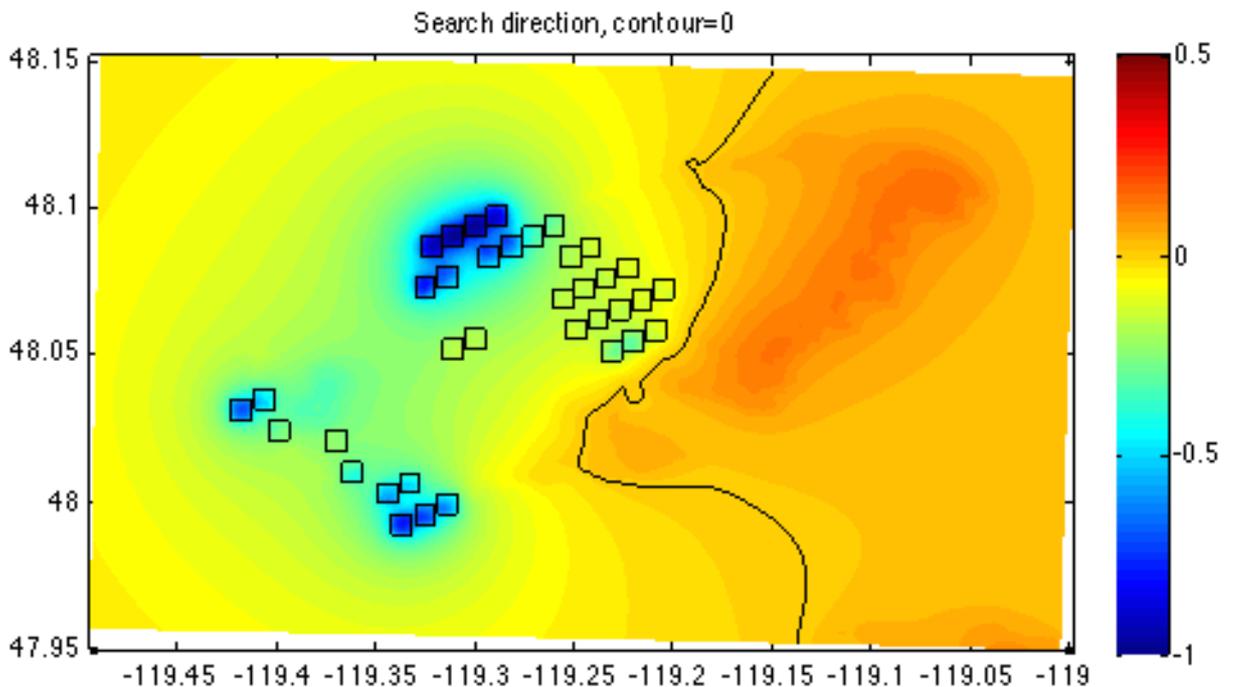

*Figure 5 – Search direction for the forecast in Fig. 3. The forecast fire arrival time will be adjusted by a positive multiple of the search direction. In particular, the fire arrival time at the location of the detection squares colored in blue may be decreased, so the fire will be there at the satellite overpass time. The black line marks the locations where the search direction is zero.*

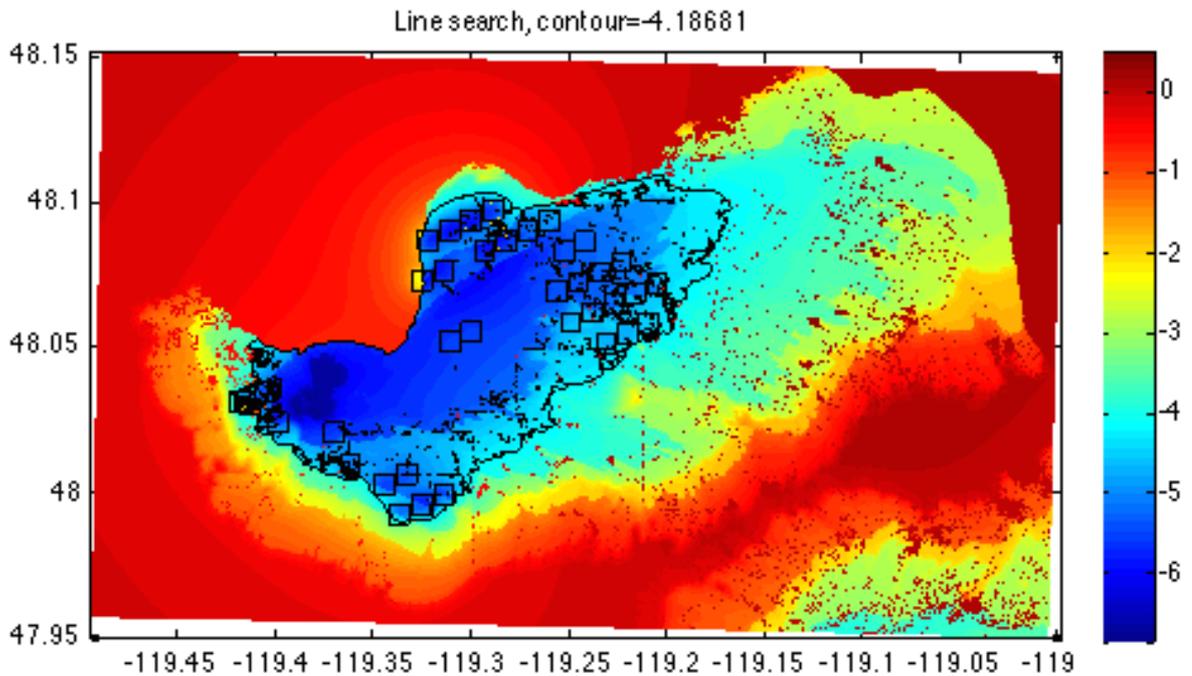

*Figure 6. Analysis for the forecast in Fig. 3. The method has filled in the fire arrival time based on the forecast from the model and incomplete information from the fire detection squares. The black contour is the resulting analysis fire perimeter.*

## 4. Conclusion

We have described a new methodology to assimilate satellite active fire detection data into the state of a fire spread model, which is encoded as the fire arrival time. The methodology relies on standard Bayesian framework of data assimilation, which leads to a nonlinear least squares problem to simultaneously minimize the change of the fire arrival time from the forecast in a suitable norm (which penalizes spatially smooth changes less) and maximize the integral of the log likelihood of the fire detection (or non-detection) over the fire simulation domain. An efficient method to solve such problems was presented, which generates smooth increments and it found good approximate solutions in a single gradient descent iteration.

There are many uncertainties affecting fire spread simulations. They are associated with limitations of the fire spread and atmospheric models as well as limited accuracy in the estimate of the initial state of the fuel and atmosphere. As the length of the atmospheric simulation grows, its accuracy deteriorates, and as a consequence, the errors in the fire spread prediction also grow. The presented method allows for objective corrections of the simulated fire progression based on the assimilation of the satellite fire detection data. As the simulation period extends, also more satellite data showing the fire progression becomes available, so the simulated fire may be cyclically nudged to the observations. The method was illustrated on two examples, in which just a single arrival time adjustment to the coupled atmosphere-fire simulations by WRF-SFIRE was made. The same method, however, can be also used for cyclic corrections of the simulated fire progression, where the model continues its fire progression from the adjusted fire state till the new observations become available. A spin-up of the atmospheric state is then needed to restart the coupled model from the modified fire arrival time. An illustration of the cyclic application of this method will be presented elsewhere.

**Acknowledgements**

This research was partially supported by the National Science Foundation (NSF) grants DMS-1216481, National Aeronautics and Space Administration (NASA) grants NNX12AQ85G and NNX13AH9G, and the Grant Agency of the Czech Republic grant 13-34856S. The authors would like to thank the Center for Computational Mathematics University of Colorado Denver for the use of the Colibri cluster, which was supported by NSF award CNS-0958354. This work partially utilized the Janus supercomputer, supported by the NSF grant CNS-0821794, the University of Colorado Boulder, University of Colorado Denver, and National Center for Atmospheric Research.